\documentclass[twocolumn,showpacs, prb,aps]{revtex4}
\usepackage{graphicx}
\begin{document}
\date{\today}
\title{Thermodynamics of
surface tension: application to  electrolyte solutions}
\author{\bf Yan Levin} 
\affiliation{\it Instituto de F\'{\i}sica, Universidade Federal
do Rio Grande do Sul\\ Caixa Postal 15051, CEP 91501-970, 
Porto Alegre, RS, Brazil\\ 
{\small levin@if.ufrgs.br}}

\begin{abstract}

In this contribution to the special issue of the Journal of
Statistical Physics dedicated to Michael Fisher on his $70$'th
birthday, I shall review two thermodynamically distinct routes
for obtaining the interfacial tension of liquid-vapor interfaces
in mixtures.  A specific application to the calculation of 
excess surface tension of aqueous electrolyte solutions will be
presented.

\end{abstract}
\maketitle
\bigskip


\section{Introduction}

It is a pleasure to dedicate this  review to Michael
Fisher on the occasion of his 70th birthday.  The 
paper deals with two subjects which are 
dear to Michael's heart,
the Thermodynamics and the Coulomb Systems.  
Here I will review two thermodynamically distinct routes to
surface tension of aqueous electrolyte solutions.  The first,
grand canonical route, goes all the way to the pioneering 
work of Gibbs on the foundations of 
thermodynamics and statistical
mechanics~\cite{gib28} and is the usual method used by physical
chemists.  Application of the Gibbs adsorption isotherm 
to the calculation of surface tension
of electrolyte solutions started with the works of Wagner~\cite{wag24}
and Onsager and Samaras~\cite{ons34} early
in the 20th century. The second, canonical method, has been
introduced recently and was found to work very well for symmetric
1:1 electrolytes~\cite{lev00,lev01}. 

\section{Grand-canonical route}

Many elementary thermodynamics and statistical
mechanics texts neglect to deal with all the 
subtleties leading to the  Gibbs adsorption
isotherm.  While these do not seriously
affect the results for surface tension 
of surfactant solutions, the
increase of  interfacial tension of water due
to salts is proportionately much smaller.
Thus, a particular care must be taken
with the thermodynamics in order to
account for all the relevant contributions.
We start, therefore, by reviewing the
thermodynamics leading to the Gibbs adsorption
isotherm~\cite{gib28,ran63}. 

Consider an aqueous solution in equilibrium with
vapor, Fig 1.
\begin{figure}  
\begin{center}
\includegraphics[width=9cm]{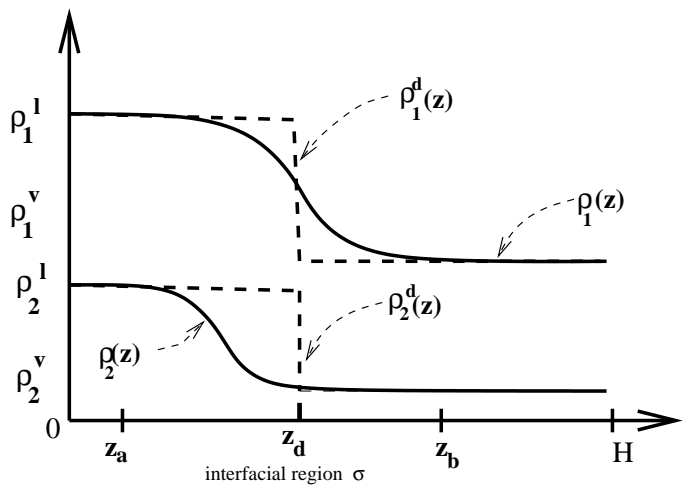}
\end{center}
\caption{Schematic density profiles inside a 
solution: $\rho_1(z)$ is
the characteristic density variation  of water across the liquid-vapor
interface, $\rho_2(z)$ is the  
concentration profile of solute. The hypothetical dividing
surface separating the liquid from the vapor is 
located at $z=z_d$.  The discontinuous ``bulk'' profiles
for solvent and solute are $\rho_1^d(z)$ and $\rho_2^d(z)$,
respectively.
Note that differently from non-ionic solutes for which $\rho_2(z)$
extends all the way into the vapor phase, because of a favorable
gain in solvation free energy aqueous electrolytes 
are completely confined to the liquid phase.}
\label{Fig1}
\end{figure}
The bulk density of 
liquid is $\rho^l_1$ and the bulk density of vapor is
$\rho^v_1$.  The variation of density is confined
to the interfacial region $\sigma$ located between
$z_a$ and $z_b$, where $z$ is the axis perpendicular to the interface.  
The thermodynamic equilibrium
requires the constancy of chemical potentials
of solute and solvent and of the pressure throughout the
system.  The differential internal energy of the interfacial region 
$\sigma$ is
\begin{equation}
\label{1}
dE^{\sigma} = TdS^\sigma - PdV^\sigma + \gamma dA^\sigma + \sum \mu_i 
dN_i^\sigma \;,
\end{equation}
where $T,P,S,V,A,\gamma, \mu_i,N_i$ 
are the temperature, pressure, entropy, volume, area, surface tension,
chemical potential, and the number of particles of type $i$. The
superscript $\sigma$ stands for the interfacial properties.
The sum runs over all the species, solute and solvent.
Since the internal energy $E$ is an extensive 
function of $\{S,V,A,N_i\}$, application of Euler theorem for
first order homogeneous functions allows the  
integration of Eq.~(\ref{1}) yielding,
\begin{equation}
\label{2}
E^{\sigma} = TS^\sigma - PV^\sigma + \gamma A^\sigma + \sum \mu_i 
N_i^\sigma \;.
\end{equation}
As usual the Gibbs free energy for the interface
is defined through the Legendre transform
\begin{equation}
\label{3}
G^{\sigma} = E^\sigma - TS^\sigma + PV^\sigma - \gamma A^\sigma \;,
\end{equation}
which after substitution of internal energy,  Eq.~(\ref{2}),
reduces to
\begin{equation}
\label{4}
G^{\sigma} = \sum \mu_i N_i^\sigma\;.
\end{equation}
The Gibbs free energy is a natural function of $\{T,P,\gamma,N_i^{\sigma}\}$
and its differential is
\begin{equation}
\label{5}
dG^{\sigma} =  -S^\sigma dT + V^\sigma dP -  A^\sigma d\gamma + \sum \mu_i 
d N_i^\sigma \;.
\end{equation}
On the other hand differentiating Eq.~(\ref{4}), we find
\begin{equation}
\label{6}
dG^{\sigma} = \sum \mu_i d N_i^\sigma + \sum N_i^\sigma d\mu_i \;.
\end{equation}
Comparing Eq.~(\ref{5}) with Eq.~(\ref{6}) we are led to a Gibbs-Duhem-like
equation for the interface,
\begin{equation}
\label{7}
S^\sigma dT - V^\sigma dP +  A^\sigma d\gamma + \sum N_i^\sigma d\mu_i = 0 \;.
\end{equation}

Now lets consider in more detail a two component system.  
In this case Eq.~(\ref{7}) simplifies to
\begin{equation}
\label{8}
S^\sigma dT- V^\sigma dP +  A^\sigma d\gamma + N_1^\sigma d\mu_1
+ N_2^\sigma d\mu_2= 0 \;.
\end{equation}
The chemical potentials are uniform throughout the system
and their variations can be obtained by considering the
bulk liquid phase.
Remembering that a change in concentration of solute 
affects the chemical potentials of both solute and solvent,
we have
\begin{equation}
\label{9}
d\mu_1=-s_1dT + v_1 dP+
\left(\frac{\partial \mu_1}{\partial c_b}\right)_{T,P} dc_b 
\end{equation}
and
\begin{equation}
\label{10}
d\mu_2=-s_2dT + v_2 dP+
\left(\frac{\partial \mu_2}{\partial c_b}\right)_{T,P} dc_b \;,
\end{equation}
where the $s_1$ and $v_1$ are the partial entropy and volume per
particle of solvent, $s_2$ and $v_2$ are the partial entropy and volume
per particle of solute, and $c_b$ is the concentration
of solute, all the values taken inside the bulk liquid phase.  
Substituting  Eqs.~(\ref{9}) and (\ref{10}) into
Eq.~(\ref{8}), the Gibbs-Duhem equation for the interface
becomes,
\begin{eqnarray}
\label{11}
&& (S^\sigma-N_1^\sigma s_1-N_2^\sigma s_2) dT  - \\ \nonumber
&& (V^\sigma-N_1^\sigma v_1-N_2^\sigma v_2) dP + \\ \nonumber
&&  A^\sigma d\gamma +
\left[ N_1^\sigma \left(\frac{\partial \mu_1}{\partial c_b}\right)_{T,P}
+ N_2^\sigma\left(\frac{\partial \mu_2}{\partial c_b}\right)_{T,P}\right] 
dc_b= 0 \;.
\end{eqnarray}
For a truly two component system pressure cannot be held constant if $T$ 
or $c_b$ is varied.  However, presence of an inert gas
does not significantly affect aqueous surface tension,
so that in practice $P$ can  be kept fixed~\cite{ran63}. 
With the help of the Gibbs-Duhem equation for the bulk liquid,
\begin{eqnarray}
\label{12}
 N_1 \left(\frac{\partial \mu_1}{\partial c_b}\right)_{T,P}
+ N_2 \left(\frac{\partial \mu_2}{\partial c_b}\right)_{T,P} 
= 0 \;,
\end{eqnarray}
where $N_1$ and $N_2$ are the bulk numbers  of solvent and solute 
molecules,  Eq.~(\ref{11}) reduces to two equations,
\begin{eqnarray}
\label{13}
\left(\frac{\partial \gamma}{\partial T}\right)_{c_b,P}&=&
-\left[\frac{S^\sigma}{A^\sigma}-\Gamma_1^\sigma s_1-\Gamma_2^\sigma s_2\right] \;,
\end{eqnarray}
and,
\begin{eqnarray}
\label{14}
\left(\frac{\partial \gamma}{\partial \mu_2}\right)_{T,P}&=&
-\left[\Gamma_2-\frac{N_2}{N_1} \Gamma_1\right] \;,
\end{eqnarray}
where $\Gamma_1=N_1^\sigma/A^\sigma$ and $\Gamma_2=N_2^\sigma/A^\sigma$.  
Equation (\ref{13}) states that the variation of surface tension
with respect to temperature is minus the excess entropy, 
compared to the entropy content  of the same amount of 
material  inside the liquid phase. The equation (\ref{14}), called
the Gibbs adsorption isotherm,  shows that the
change of surface tension with respect to chemical potential of solute
is minus the excess of solute inside $\sigma$,  over 
the amount which would be present for the same
quantity of solvent in a bulk liquid phase. 
Note that the term ``excess'' is
used both when the interfacial region  has higher (positive excess) or 
lower (negative excess) 
concentration of solute, as compared to the bulk liquid phase.  
Since the thermodynamic stability 
requires $\partial \mu_2 / \partial c_b > 0$ at fixed temperature, 
Eq.~(\ref{14}) shows that
a positive surface
excess of solute leads to a lower interfacial tension,
while a negative surface excess of solute results 
in a higher surface tension.
Finally, we note that the  Eq.~(\ref{14})
is invariant with respect to the specific location 
of the dividing surfaces $z_a$
and $z_b$, as long as they completely enclose the 
region of strong density variation.

The surface
tension can be obtained by integrating Eqs.~(\ref{13}) and (\ref{14}). 
Of course, this requires a specific microscopic model which would
allow us to calculate the excess entropy and the excess amount of 
solute inside $\sigma$. Such model can, in principle, 
be provided by statistical mechanics~\cite{ons34,sch55}.

In the language of statistical mechanics the calculations based
on Eqs.(\ref{13}) and (\ref{14}) are intrinsically grand-canonical.
The number of particles inside  $\sigma$ fluctuates and is determined
by the condition of equilibrium between 
the interfacial region and the bulk phases.
Unfortunately to actually solve a statistical mechanical model,
one is invariably forced to make  approximations.
It might be, therefore,  worthwhile to explore different routes
to surface tension, each one requiring different approximations.
Of course, in an exact calculation all the thermodynamic 
routes will lead to the same
result. With approximate theories this, however, is no longer 
the case and some routes can be significantly better than others. 
With this in sight, we have explored the canonical route to surface 
tension~\cite{lev00,lev01}.

\section{Canonical route}

Suppose that the system depicted in Fig. 1 
which contains 
\begin{equation}
\label{14a}
N_1 = A\int_0^{H} \rho_1(z) dz
\end{equation}
particles of solvent and 
\begin{equation}
\label{14b}
N_2 = A\int_0^{H} \rho_2(z) dz
\end{equation}
particles of solute is confined to a cylindrical 
box of height $H$ and cross-sectional area $A$.
We shall define the ``bulk'' density of solvent and solute inside the 
liquid phase as $\rho_1^l=\rho_1(0)$ and $\rho_2^l=\rho_2(0)$, respectively; 
and the ``bulk'' density
of solvent and solute inside the vapor phase as $\rho_1^v=\rho_1(H)$
and $\rho_2^v=\rho_2(H)$. 
The internal energy of the whole system is
\begin{equation}
\label{15}
E = TS - PV + \gamma A + \mu_1 N_1+ \mu_2 N_2 \;,
\end{equation}
Now we would like to ask which part of this energy is due to  
the interface?  That is, if instead of a continuous density
profiles $\rho_1(z)$ and  $\rho_2(z)$, 
we would have two discontinuous ``bulk'' 
profiles $\rho_1^d(z)$ and  $\rho_2^d(z)$,
depicted in
Fig. 1, what would be the change in internal energy of the
system?  Although easy to pose, this question carries some
subtlety.  Specifically where should we put the dividing surface?
Also, what is the meaning of a discontinuous profile?  This latter
question is fairly easy to answer. To obtain
a discontinuous profile we divide the total volume 
into two sub-cylinders
of heights $z_d$ and $H-z_d$.
In the first sub-cylinder of height $z_d$  and volume  $V^l=A z_d$, 
there will be  
$N_1^l(z_d)=\rho_1^l V^l$ molecules of solvent and  
$N_2^l(z_d)=\rho_2^l V^l$ molecules of solute. 
In the second sub-cylinder of height $H-z_d$ and volume $V^v=A (H- z_d)$,
there will be  $N_1^v(z_d)=\rho_1^v V^v$ molecules of solvent and
$N_2^v(z_d)=\rho_2^v V^v$ molecules of solute. 
To keep the uniform density distribution characteristic
of the bulk phases, we impose  periodic boundary conditions in the 
$z$ direction on each sub-cylinder. 
The internal energy of the first sub-cylinder, 
corresponding to the bulk liquid,
is then
\begin{equation}
\label{18}
E^l(z_d) = TS^l(z_d) - P V^l + \mu_1 N_1^l(z_d)+ \mu_2 N_2^l(z_d) \;,
\end{equation}
and the internal energy of the second sub-cylinder, corresponding to the
bulk vapor, is
\begin{equation}
\label{19}
E^v(z_d) = TS^v(z_d) - P V^v + \mu_1 N_1^v(z_d)+ \mu_2 N_2^v(z_d) \;.
\end{equation}
Because of the arbitrariness in the location of
the dividing surface, in general, $N_1 \neq N_1^l(z_d)+N_1^v(z_d)$ and 
$N_2 \neq N_2^l(z_d) + N_2^v(z_d)$. Therefore the 
surface internal energy, $E^s(z_d)=E-E^l(z_d)-E^v(z_d)$, is
\begin{equation}
\label{20}
E^s(z_d) = TS^s(z_d) + \gamma A + \mu_1 N_1^s(z_d)+ \mu_2 N_2^s(z_d) \;,
\end{equation}
where 
$S^s(z_d)=S-S^l(z_d)-S^v(z_d)$, 
$N_1^s=N_1-N_1^l(z_d)-N_1^v(z_d)$, and  
$N_2^s=N_2-N_2^l(z_d)-N_2^v(z_d)$.
Since the Helmholtz free energy is given in terms of the Legendre
transform of the internal energy, $F^s=E^s-TS^s$, Eq.~(\ref{20})
can be rewritten as,
\begin{equation}
\label{21}
F^s(z_d)-\mu_1 N_1^s(z_d)-\mu_2 N_2^s(z_d)=\gamma A  \;.
\end{equation}

The statistical mechanics allows us, in principle, to 
calculate the excess surface Helmholtz free energy as well as
the chemical potentials and the  surface excess of
solvent and solute.  The thermodynamics relates these to 
surface tension through Eq.~(\ref{21}).  Note that
while the three terms on the 
left hand side of Eq.~(\ref{21}) are dependent on the
location of the dividing surface, the right hand
side does not.  Thus, we can
fix the position of $z_d$ so that the surface
excess of solvent is zero, $N_1^s(z_d^G)=0$. This specific choice
corresponds to the, so called, Gibbs dividing surface. With the location
of $z_d^G$ specified, the surface
tension becomes
\begin{equation}
\label{22}
\gamma = \frac{F^s(z_d^G)-\mu_2 N_2^s(z_d^G)}{A}  \;,
\end{equation}
Now, lets  define the {\it liquid} surface excess of solute as
\begin{equation}
\label{22b}
\Delta^l=A\int_0^{z_d^G} [\rho_2(z)-\rho_2^l] dz\;,
\end{equation}
and the  {\it vapor} surface excess of solute as
\begin{equation}
\label{22c}
\Delta^v=A\int_{z_d^G}^H [\rho_2(z)-\rho_2^v] dz\;.
\end{equation}
Noting that $N_2^s=\Delta^l+\Delta^v$
and
\begin{equation}
\label{22a}
\mu_2=\frac{\partial F_{bulk}^l}{\partial N_2^l} \Big\arrowvert_{T,V^l}=
\frac{\partial F_{bulk}^v}{\partial N_2^v} \Big\arrowvert_{T,V^v}\;,
\end{equation}
Eq.~(\ref{22}) can be rewritten as
\begin{eqnarray}
\label{22d}
\gamma = \frac{1}{A} \left[F(N_1,N_2)-F_{bulk}^l(N_1^l,N_2^l)-   
F_{bulk}^v(N_1^v,N_2^v)-\right. \nonumber \\ 
\frac{\partial F_{bulk}^l }{\partial N_2^l} \Delta^l
 \left. -  \frac{\partial F_{bulk}^v }{\partial N_2^v} \Delta^v \right]  \;,
\end{eqnarray}
where $F(N_1,N_2)$ is the total Helmholtz free of the system with interface.
 
In the thermodynamic limit $\Delta^l/N_2^l \ll 1$ and $\Delta^v/N_2^v \ll 1$,
and  Eq.(\ref{22d}) simplifies to 
\begin{eqnarray}
\label{22e}
\gamma = \frac{1}{A}[F(N_1,N_2)-F_{bulk}^l(N_1^l,N_2^l+\Delta^l)- \\ \nonumber   
F_{bulk}^v(N_1^v,N_2^v+\Delta^v)]  \;.
\end{eqnarray}

This is the principal thermodynamic result of this  paper, which
will serve as the starting point for our analysis of the interfacial
tension of electrolyte solutions.
The fact that $N_1=N_1^l+N_1^v$ and 
$N_2=N_2^l+N_2^v+\Delta^l+\Delta^v$
signifies that the calculations based on Eq.~(\ref{22e}) must
be performed with a {\it fixed} number
of solvent and solute molecules. 
This accounts for the adjective ``canonical''
in the name of this thermodynamic 
route to surface tension.

For strong electrolytes, 
we have an additional simplification. Because of the
high dielectric constant of water, ions inside the liquid  have a
significantly lower electrostatic free energy compared to ions
inside the vapor.  The probability of finding an
ion inside bulk liquid $p_l$, compared to its probability of 
being inside bulk vapor  $p_v$ is
\begin{equation}
\label{22f}
\frac{p_l}{p_v} = e^{-\beta W}\;,
\end{equation}
where $\beta=1/k_B T$ and $W$ is the solvation energy, 
which can be estimated from
the Born equation,
\begin{equation}
\label{23}
W=\frac{q^2}{d} \left(\frac{1}{\epsilon_l}-\frac{1}{\epsilon_v}\right)\;.
\end{equation}
In this formula $q$ is the ionic charge, $d$ is the ionic
diameter, $\epsilon_l$ is the dielectric constant of liquid,
and  $\epsilon_v$ is the dielectric constant of vapor. For
water $\epsilon_l/\epsilon_v \approx 80$, so that $\beta W$
can be well approximated by $-\lambda_B \epsilon_l/d\epsilon_v$, 
where $\lambda_B=\beta q^2/\epsilon_l$ is 
the Bjerrum length.  For water at room 
temperature and monovalent ions $\lambda_B =7.2$ \AA,  while the
characteristic diameter of a hydrated ion is approximately 
$d \approx 4$ \AA.
Therefore, the concentration of ions in  vapor can be estimated
to be $\rho_2^v=\rho_2^l \exp(-140)$, which   
is immeasurably low. 
Furthermore, a rapid decrease of the dielectric constant 
for $z>z_d^G$ should strongly inhibit presence of ions
in the interfacial region,  confining them 
to $z<z_d^G$, so that $N_2^v=0$ and $\Delta^v=0$. 

\section{A simple model of an electrolyte}

Let us now consider a simple model of an aqueous 
electrolyte~\cite{lev01} 
confined
to a cylinder of cross-sectional area $A$ and height $H$. The $N$ ions
will be idealized as hard spheres of diameters $d$,
carrying charge $+q$ or $-q$ at their center. Suppose
that the Gibbs dividing surface is located at the top of the
cylinder. To simplify the model, we shall further 
assume that on crossing the Gibbs dividing surface from bellow 
the dielectric constant drops discontinuously 
from $\epsilon=80$, characteristic of bulk water, 
to $\epsilon=1$, characteristic of vacuum.
From the previous  discussion, 
the increase in  surface tension of
water due to electrolyte is
\begin{equation}
\label{24}
\gamma^{ex} =\lim_{A,H,N \rightarrow \infty} \frac{1}{A} 
(F^{ex}-F^{ex}_{bulk})  \;,
\end{equation}
where the thermodynamic limit is taken in such a way as to
preserve the bulk concentration, $c_b=N/2 A H$. In Eq.~(\ref{24}),  
$F^{ex}$ is the excess Helmholtz free energy due to
electrolyte in the presence of liquid-vapor
interface,  while $F^{ex}_{bulk}$ is
the excess free energy of the {\it same amount} of electrolyte but dissolved 
within  the bulk liquid phase of the same volume. 
The  $F^{ex}_{bulk}$ can be calculated using the
periodic boundary conditions at the top and the bottom of the
cylinder.

As was already mentioned, the decrease in free energy  
due to solvation inhibits entrance of ions into the interfacial
region, producing an  ion-free layer adjacent to
the Gibbs dividing surface.  The different 
hydration characteristics of anions and cations can make their
exclusion layers to have different width.  Let us call the width
of a cation exclusion layer  $\delta_+$ and the width of an
anion exclusion layer $\delta_-$.  Furthermore, lets for the moment
neglect all the electrostatic and hardcore interactions, beyond their
contribution to the formation of ion-free layers. In this case
the total free energy of an electrolyte solution inside the cylinder 
takes a particularly simple entropic form,
\begin{equation}
\label{25}
F^{ex}=k_B T N_+\left[\ln(c_+ \Lambda^3)-1\right]+
k_B T N_-\left[\ln(c_- \Lambda^3)-1\right]  \;,
\end{equation}
where $N_+=N_-=N/2$, $c_+=N_+/A(H-\delta_+)$,  
$c_-=N_-/A(H-\delta_-)$, and $\Lambda$ is the thermal de Broglie wavelength. 
On the other hand the bulk 
free energy, $F^{ex}_{bulk}$,
is given by exactly the same expression, but
with $c_+=c_-=c_b \equiv N/2AH$, 
since the  periodic boundary condition destroys 
the ion free layers. Substituting into Eq.~(\ref{24}) we find
a simple result,
\begin{equation}
\label{26}
\gamma^{ex}=k_B T c_b (\delta_++\delta_-)  \;.
\end{equation}
Neglect of direct electrostatic interactions 
results in  excess surface tension
scaling linearly with the concentration of 
electrolyte~\cite{sch55,ran63}.  Eq.~(\ref{26}) is quite
different from the Onsager-Samaras ($OS$) limiting
law~\cite{ons34},
\begin{equation}
\label{27}
\gamma^{ex}_{LL}=\gamma_0[-\ln(\kappa \lambda_B/2)-2\gamma_E+3/2]  \;.
\end{equation}
where $\gamma_E=0.577215665...$ 
is the Euler's constant, $\gamma_0=q^2 c_b/2 \epsilon $, 
and  $\kappa= \sqrt{8 \pi q^2 c_b/\epsilon k_B T}$ is 
the inverse Debye length. The $OS$ limiting law is believed to be 
universally valid for all electrolytes at infinite dilution. In this
respect, it is supposed to be analogous to the Debye limiting laws for
bulk electrolytes.  Comparing Eq.~(\ref {27}) with
Eq.~(\ref {26}) we see, however, that only the first term inside the square
brackets is universal, while the existence of ion free layers 
leads to corrections which scale linearly with the
electrolyte concentration. Although,  Onsager and Samaras
tried to extend their theory to finite densities, they have
overlooked existence of ion free layers and thus missed
an important contribution to the surface tension.   

The experimental measurements show a purely linear
dependence of excess surface tension on concentration~\cite{mat98}, 
with the validity of $OS$ limiting law restricted to very low electrolyte
densities.  Neglect of electrostatic interaction is clearly
a strong oversimplification. The advantage
of the canonical formalism, besides its thermodynamic
simplicity, is that the electrostatics can be easily 
taken into account.
Thus, the Debye-H\"uckel theory~\cite{deb23,fis93,lev96} for 
bulk electrolytes
can be  extended to take into account the liquid-vapor 
interface~\cite{lev00,lev01}. 
This calculation, which is in excellent agreement with experiments,
shows that for $NaCl$, existence
of an ion free layer of width $\delta_+=\delta_-=2.125$ \AA,
characteristic of ionic hydration radius,
accounts for about $30 \%$ of the total contribution
to $\gamma^{ex}$, for $0.5 M$  to $1 M$ solutions. 
Electrostatics becomes even more important at lower 
concentrations.  For solutions with $c_b<0.3 M$ the electrostatics
is  over $80 \%$ dominant, but the  
$OS$ limiting law fails to be a reasonable approximation until 
$c_b<0.1 M$. Furthermore, even an extended grand-canonical 
calculation~\cite{ons34} which goes beyond the limiting laws, 
significantly underestimates the role of 
electrostatics~\cite{lev00,lev01}.
This suggests that the canonical route to surface tension of
aqueous electrolytes is more reliable.

It would be interesting to see if the electrostatic calculations
for symmetric electrolytes can be extended to the asymmetric
systems with distinct $\delta_+$ and $\delta_-$, perhaps along the same
lines as used recently by Michael Fisher and collaborators 
for the asymmetric bulk electrolytes~\cite{zuc01}.

\section{Acknowledgement}
It is a
pleasure to acknowledge interesting conversations with A. Parsegian.
This work was supported in part by  
Conselho Nacional de
Desenvolvimento Cient{\'\i}fico e Tecnol{\'o}gico (CNPq).

\end{document}